\documentclass{article}
\usepackage{amsmath}
\usepackage{comment}
\usepackage[font=footnotesize, width=1.1\textwidth, labelfont=bf]{caption}

\usepackage[linesnumbered,lined,ruled,longend]{algorithm2e}

\let\oldnl\nl
\newcommand{\nonl}{\renewcommand{\nl}{\let\nl\oldnl}}

\usepackage[margin=1in]{geometry}
\DeclareMathOperator*{\argmax}{argmax}

\usepackage{graphicx}

\providecommand{\keywords}[1]
{
	\small	
	\textbf{\textit{Keywords---}} #1
}

\title{Towards a Systematic Computational Framework \\ for Modeling Multi-Agent Decision-Making at Micro Level \\ for Smart Vehicles in a Smart World}
\author{Qi Dai\thanks{Corresponding Author. Telephone: 313-236-7640.}\,, \,Xunnong Xu,\,Wen Guo,\, Suzhou Huang,\, Dimitar Filev \\ 
\{qdai2,\,xxu63,\,wguo16,\,shuang10,\,dfilev\}@ford.com \\ 
Ford Motor Company, Dearborn, MI 48126}
\begin{document}

\maketitle
 
\begin{abstract}
	We propose a multi-agent based computational framework for modeling decision-making and strategic interaction at micro level for smart vehicles in a smart world. The concepts of Markov game and best response dynamics are heavily leveraged. Our aim is to make the framework conceptually sound and computationally practical for a range of realistic applications, including micro path planning for autonomous vehicles. To this end, we first convert the would-be stochastic game problem into a closely related deterministic one by introducing risk premium in the utility function for each individual agent. We show how the sub-game perfect Nash equilibrium of the simplified deterministic game can be solved by an algorithm based on best response dynamics. In order to better model human driving behaviors with bounded rationality, we seek to further simplify the solution concept by replacing the Nash equilibrium condition with a heuristic and adaptive optimization with finite look-ahead anticipation. In addition, the algorithm corresponding to the new solution concept drastically improves the computational efficiency. To demonstrate how our approach can be applied to realistic traffic settings, we conduct a simulation experiment: to derive merging and yielding behaviors on a double-lane highway with an unexpected barrier. Despite assumption differences involved in the two solution concepts, the derived numerical solutions show that the endogenized driving behaviors are very similar. We also briefly comment on how the proposed framework can be further extended in a number of directions in our forthcoming work, such as behavioral calibration using real traffic video data, computational mechanism design for traffic policy optimization, and so on.
\end{abstract} 


\newpage
\section{\bf Introduction}
Aiming to improve customer experience and traffic flow, reduce emission and congestion, vast attention and resources have been devoted to make motor vehicles autonomous and electrified. In this endeavor the main focus in the literature so far is mostly following a strategy that is vehicle centric: to make the vehicle as smart as possible, while treating the environment as a given background. As AI and communication/network technology rapidly progress, vehicles are becoming increasingly connected as well as smarter. Concurrently, infrastructure is also becoming more and more intelligent. Time appears to be ripe that we start to shift attention from the vehicle centric paradigm to a system centric one. We anticipate that a smart transportation system that will work well in reality has to explicitly address the issue of interaction and coordination among smart vehicles and the smart infrastructure, and likely other players, such as pedestrians, bicycles, scooters, and even animals. It is in this system perspective that we propose our modeling framework for smart vehicles in a smart world. Recently, we have also witnessed increasing efforts on how to take advantage of the forthcoming connectivity and autonomy, such as reducing the congestion and emission. For example, an optimal control framework \cite{Cassandras17} is put forward to address issues related to intelligent vehicles in smart city settings. 

In our approach, how to endogenize the decision-making at micro level with interaction and coordination among all players is central to the modeling. We further hope that the modeling framework can be used to quantitatively characterize human behaviors in various realistic traffic settings, and to help refine existing traffic rules or derive new ones from the perspective of the transportation authority. It turns out that such a modeling framework has already been around for quite some time: game theory. For those who are versed in microeconomics, it is apparent that the framework we are going to propose is essentially identical to what had been developed in microeconomics. This should not be a big surprise, since they both share the same theme of modeling smart agents interacting with one another within a regulated regime. The high level concepts, methodologies, and even some of the detailed mathematics are inevitably common.

There are generally three classes of questions the modeling framework will be able to address conceptually. The first one has to do with endogenizing the decision-making using optimization given preferences and game rules. This is what we call the forward problem or utility maximization in microeconomics. Here we expect that driving behaviors, such as coasting, distance keeping, yielding, merging, lane-changing, passing, entering and exiting, can be solved in various situations given physical constraints and initial conditions. One of the ultimate goals we have in mind is to provide driving instructions to autonomous vehicles at sub-second frequency in real time, taking into account multi-vehicle interactions in mixed settings where autonomous vehicles and human driven vehicles co-exist. The second class has to do with how to derive and quantify human driving behavioral models given the observed action sequences in real traffic situations. This is what we call the inverse problem, whose methodologies involved are similar to micro-econometrics and is also known as imitation learning in the machine learning community. Here we expect to extract driving preference quantitatively, such as the preferred speed on a specific road, head distance when following other vehicles, and how these depend on weather, time of the day, road condition and lighting, traffic density, and so on, by using traffic video data with explicit heterogeneity. The third class has to do with how the traffic rules can be optimized to benefit the society as a whole, and how new traffic rules can be invented to accommodate the emerging phenomena, once we understand the decision-making and preference for all players involved. This is what we call the mechanism design problem, in which we hope the modeling framework will also be helpful for dealing with potential trolley-type of problems and associated moral ambiguities that can arise unavoidably.

The other focus in our modeling framework is to make the numerical algorithms practical for a large number of realistic path planning applications, and even in real-time applications ultimately. This requires us to identify an appropriate mathematical framework that is both conceptually clean and computationally efficient. Deterministic Markov games appear to satisfy these criteria. However, realistic traffic settings are generally stochastic, and hence we introduce risk premium to ease the burden of tracking state variables precisely in the game, in the spirit of reward shaping \cite{RewardShaping}. We explore two types of solution concepts. The first one is the standard sub-game perfect Nash equilibrium in a dynamic setting. The advantage of Nash equilibrium is its conceptual and mathematical cleanness. However, this cleanness is at the expense of practical relevance due to very strong rationality assumptions and heavy computational requirements. This leads us to consider a heuristics based solution concept, i.e., adaptive optimization with finite look-ahead anticipation. We obtain this new solution concept by systematically relaxing some of the less realistic assumptions mandated by Nash equilibrium. Along the way these conceptual simplifications also open the avenue for drastically reducing the computational burden.

There is a large body of literature on applications of game theory to traffic behavioral modeling. A thorough review of them all is practically impossible. We concentrate only on those that are the closest to our approach: using game theory to address micro level decision-making with explicit interaction and coordination among multiple agents with some generality. In a series of papers, Yoo and Langari \cite{Yoo12}, Kim and Langari \cite{Kim14}, Yu {\it et al} \cite{Yu18}, Zhang {\it et al} \cite{Zhang18}, and Zhang {\it et al} \cite{Zhang20} used Stackelberg solution concept to model highway driving. In this approach strategic decisions are made sequentially, with vehicles divided into a leader and follower(s). Payoffs are given typically by a matrix form with discrete moves. Some additional technical simplifications are made to make the approach tractable. Consequently, the modeling results are mostly meant to be qualitative, and hence cannot be used for micro path planning. In a different strand of inquiry, Oyler {\it et al} \cite{Oyler16}, Tian {\it et al} \cite{tian2018adaptive}, and Li {\it et al} \cite{li2019game} utilized a hierarchical cognitive decision-making strategy, dubbed level-$k$ game theory, that is essentially one step further than Stackelberg solution concept by iterating the leader-follower sequence $k$ times. They tackled settings of highway driving and un-signaled intersections. Interesting results were obtained, though some additional assumptions or technical simplifications were made, again for tractability. It is not obvious that their approach can be applied for micro path planning and how to prevent occasional accidents from happening. 

Lane changing behaviors were modeled using game theory by a number of authors. Talebpour {\it et al} \cite{Talebpour15} also used a leader-follower matrix game with discrete moves to model discretionary lane-changes. They even calibrated their model using NGSIM data \cite{NGSIM2006}. They further claimed that their lane-change model provided a greater level of realism than a basic gap-acceptance model. Instead of sequential game, Meng {\it et al} \cite{Meng16} used a discrete simultaneous-move matrix game with receding horizon adaptively applied to model highway lane changes. One important feature of their approach is its use of reachability analysis to guarantee the safety. Because of the game only involves discrete moves it also cannot be used for micro path planning. In a more systematic approach, the adaptive receding horizon technique is also deployed in a differential cooperative game for predictive lane-changing and car-following control by Wang {\it et al} \cite{Wang15}. A decomposition technique is used to divide the original problem into smaller pieces and an iterative algorithm based on Pontryagan's maximum principle is then applied to solve each sub-problem. However, the lane change in this paper appears to be instantaneous, which in turn limits its applicability. Motivated by the computational efficiency's perspective, Huang {\it et al} \cite{huang2019game} introduced a radically different approach: a mean field to represent the average behavior of all agents in a local vicinity. When the number of agents $N$ becomes large, the pairwise interaction is supposed to be ignorable, and the problem is vastly simplified. The original multi-agent game is then approximated by the mean field solution to be within the realm of $\epsilon$-equilibrium. While theoretically interesting, the condition of $N$ being large in a physical world is hard to meet, especially when lateral movements as well as longitudinal movements are modeled simultaneously.

In a totally different modeling strategy, Cunningham {\it et al} \cite{Cunningham15}, Galceran {\it et al} \cite{Galceran17}, and Mehta {\it et al} \cite{Mehta18} utilized the so-called multi-policy decision-making framework. In this framework, while a decision-maker is contemplating its own move, a ``library'' of policies (either discrete or continuously parameterized) is evaluated, via simulation, for anticipating potential future moves that can be dangerous. They also proposed a gradient-based algorithm to improve the policy parameterization online. This type of approaches appears to be very promising. It will be interesting to see how this type of approach can be successfully applied to micro level modeling of traffic behaviors. One challenge seems to be in having good enough handcrafted starting policies from which the algorithm can figure out ways to refine themselves. Lastly, there is also a relatively big literature in using centralized decision-making to improve traffic systems, taking advantage of the emerging smart environment. Techniques utilized can be either multi-agent based simulation, such as Dresner and Stone \cite{dresner2008multiagent}, or based on cooperative game theory, such as Elhenawy {\it et al} \cite{elhenawy2015intersection} and Ding {\it et al} \cite{ding2017centralized}. While these papers studied the situations at automated intersections, Rios-Torres and Malikopoulos \cite{rios2017automated} investigated vehicle merging situations at highway on-ramps. For the latter case, even game engine was utilized in a recent work of Wang {\it et al} \cite{wang2018agent}.

Of course, there are plenty of research on micro path planning based on feasibility in autonomous driving literature. In this strand of endeavor surrounding agents are often treated as passive background and hence their strategic behavior and coordination are not explicitly endogenized. A recent review of this type of approaches can be found in Katrakazas {\it et al} \cite{KATRAKAZAS15}.

Our work distinguishes from these previous work in several important ways. We first introduced risk premium into the utility function so that the burden of tracking state evolution precisely is substantially reduced. Consequently, we can treat the traffic behavioral modeling as deterministic Markov games. When we develop our game theoretical solution concept in Section \ref{SolutionConcept1} ({\it betaNash}) we actually provided an iterative algorithm, based on best response dynamics, that can be used to solve sub-game perfect dynamic Markov Nash equilibrium. In our approach, no additional artificial assumptions are made beyond those required by the standard non-cooperative game theory. The solution is sufficiently detailed and can be used for micro path planning. In an attempt to relax the common strong assumptions associated with game theory so that the behavioral modeling can be directly applied to human drivers, and to substantially improve the computational efficiency, we introduced an adaptive optimization algorithm with finite look-ahead anticipation in Section \ref{SolutionConcept2} ({\it adaptiveSeek}). This new solution concept is systematically simplified from the game theory solution concept guided by human driving heuristics.  Even though some of the important modeling ingredients had appeared in some of the earlier papers cited above, such as receding time horizon, anticipation, and adaptivity, our starting point and detailed implementation are quite different. We further show, despite the conceptual difference for the two solution concepts, one being game theory based and the other being almost unilateral optimization based, the numerical results of the solutions are reasonably close. Lastly, we also try to be very careful on treating the information in the solution concepts: what is shared as the common knowledge and what is not as private intentions.

Incidentally, one recent article \cite{Schwarting19} deserves special mention here. This is the work that comes closest to ours, as far as the computational Nash equilibrium algorithm is concerned. Yet, their work is motivated very differently from ours, to embed social psychology into game theory so as to better predict driver behavior. In contrast we are motivated to provide a realistic computational framework for micro path planning. The detailed formulations of the utility function and the associated interpretations are also quite different. Their emphasis is in the heterogeneity of Social Value Orientation (SVO), whereas ours is in introducing risk premium in the pairwise collision of the utility so that we can convert a would-be uncertain game model into deterministic one, and hence reducing the computational burden. Nevertheless, their computing methodology is essentially identical to our {\it betaNash}, both are based on the concept of iterative best response, though our algorithm {\it adaptiveSeek} is entirely novel. In addition, they further analyzed the NGSIM data \cite{NGSIM2006} to quantify SVO at individual driver level by combining the Nash equilibrium algorithm with Max Entropy Inverse Reinforcement Learning \cite{MaxEntropyIRL}. It is our belief that, despite of the improved fitting, SVO as an individual's preference parameter, should not be made to vary at a time scale of second. If one driver is an altruist, he/she should not suddenly become an egoist a second later. It is much more preferable conceptually that the fast changing role should be taken up by the action variables instead. On the other hand, we will describe our approach to the inverse problem, calibrating utility function with observed action data, using state space model in a separate paper \cite{CalibrationPaper}. The utility parameters in our case, while being heterogeneous individually, are all constant during the course of 250 seconds of driving in Sugiyama experiment \cite{Sugiyama08}.

Finally, it is also worthwhile to compare our approach with approaches utilizing some form of deep reinforcement learning (RL), such as those in \cite{Shalev-Shwartz2016, Kuefler2017, Zhang2018, Hoel2018, Nageshrao2019}. One of the common characteristics of RL approaches is that they are all policy oriented, in the sense that optimal policies are typically trained offline and then deployed/generalized in similar contexts. Most of the times, the offline training is done in a manner of exploration/exploitation, a strategy that is often very slow. The heavy computational burden implies that inverse problems and mechanism design problems become very hard to tackle under this framework. Furthermore, since it embodies the specific information of the training context, the trained optimal policy can have hard time to generalize to new contexts. In contrast, our approach is objective oriented, in the sense that the optimal (or equilibrium) policy is solved for a given context online with computational efficient algorithms. Consequently, issues in inverse problems and mechanism design problems, and generalizability are all become easier (see Section \ref{SolutionConcept2} for further comments).  

The rest of this paper is organized as follows. In the next section we detail the modeling environment and state the relevant assumptions. Markov games is proposed to be the mathematical framework. We then explore two solutions concepts in the two subsequent sections. The first solution concept is based on sub-game perfect Nash equilibrium, as described in Section \ref{SolutionConcept1}. We exploit the best response dynamics to numerically solve the equilibrium resulting Algorithm {\it betaNash}. The second solution concept is based on heuristic adaptive optimization that is obtained by systematically relaxing some of the less realistic assumptions associated with Nash equilibrium, as described in Section \ref{SolutionConcept2}. The simplified but more realistic approach is dubbed Algorithm {\it adaptiveSeek}. We further show how noise/uncertainty can be accommodated within the framework, and outline how it can be utilized to formulate inverse problem using state space model approach. To demonstrate how the framework can be applied to tackle concrete problems, we consider a simulation experiment in Section \ref{SimulationExperiment}. In this experiment two vehicles in a setting of a double-lane highway with an unexpected barrier are studied in details. Solutions derived by using {\it betaNash} and {\it adaptiveSeek} are compared and contrasted. In the final section we summarize our achievements and highlight forthcoming work.

\section{\bf Modeling Environment and Assumptions}
In this section we first establish the physical assumptions that are reasonable for modeling smart vehicles in a smart world. Following the approach of microeconomics, we also outline the high level conceptual modeling framework. Then, we detail the mathematical considerations to make our modeling framework concrete and practical. 

\subsection{\bf Assumptions}
First, the agents involved can be classified into the following types:
\begin{itemize}
\item The transportation authority is a special agent who has the power to decide on the rules of the game, such as the right-of-the-way, traffic infrastructure and signal systems. Its decision-making is governed by some proxies of social aggregate outcomes. In certain contexts, the transportation authority can also serve as the controller for centralized automated transportation systems.
\item Human agent (excluding the transportation authority) is regarded as rational whose decision-making is governed by maximizing an intrinsic utility function which potentially can be calibrated by actually observed behaviors. Examples are human driven vehicles, bicycles, and pedestrians.
\item Algorithmic agent is also regarded as rational whose decision-making is governed by maximizing a utility function which needs to be {\it endowed} according to certain rules. Examples are autonomous vehicles, traffic signals or infrastructure facilities.
\item Other agents include unexpected debris, and animals that are possibly irrational and hence treated as random events.
\end{itemize}
Second, in the settings of smart vehicles in a smart world, the following simplifying assumptions appear to be appropriate:
\begin{itemize}
\item The transportation authority always has strategic precedence, relative to all other agents in the game. This is equivalent to saying that the transportation authority is the first mover of the traffic behavior game, as we assume that the transportation authority sets the rules of the game beforehand.
\item Vision, sensing, mapping, and awareness of the surroundings are adequately provided to all agents, with possible latency and communication interruption, either by the smart vehicles or the edge-based smart infrastructure. However, we do not generally assume common knowledge on everything, such as individual's driving preference, destination, and so on, as this may be private, unless an explicit communication mechanism is specified.
\item Agents are sufficiently smart so that they, when required, can faithfully execute the decision recommendations provided from the system, within the scope of the physical constraints. This implies that we can concentrate on high level variables, such as locations, velocities, accelerations, orientations, rather than worry about gas pedals, steering wheels, voltages, or actuators.
\end{itemize}

\begin{figure}[h]
\centering
\includegraphics[width=.80\textwidth]{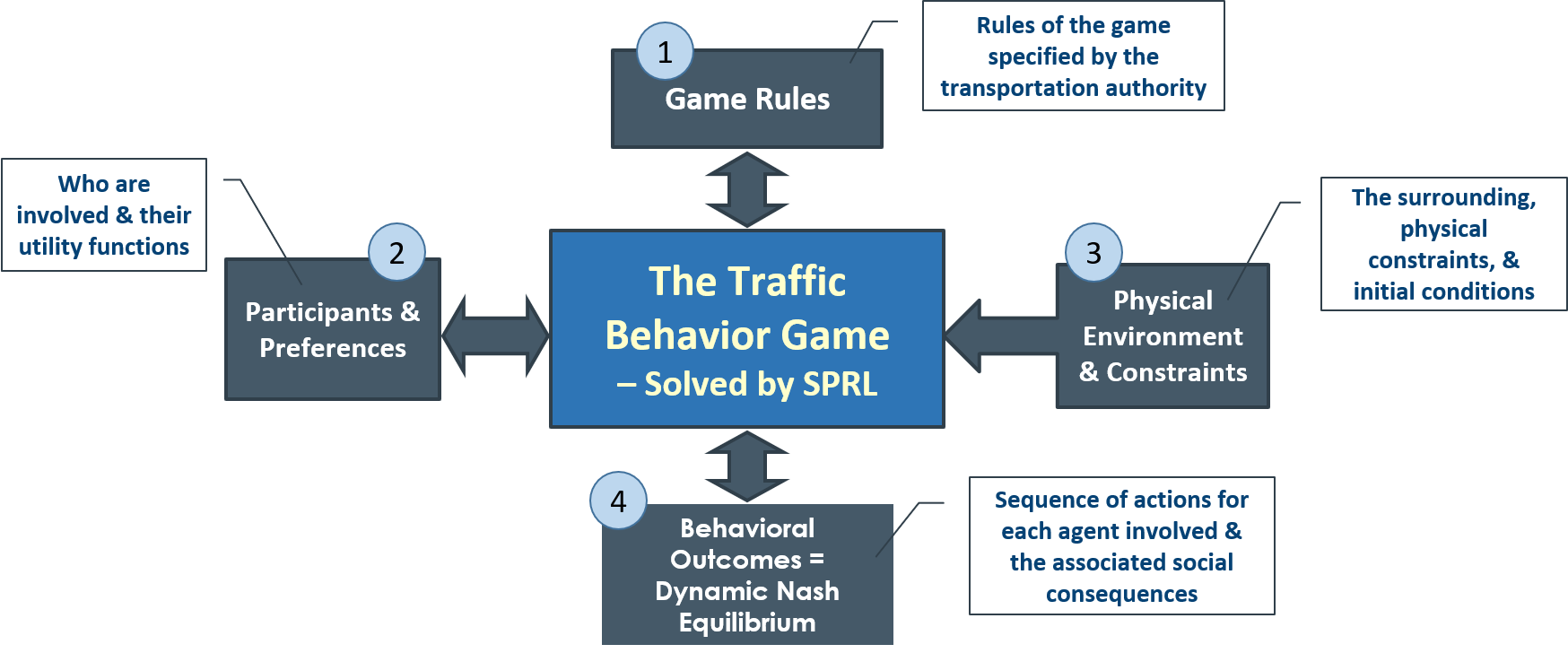}
\caption{An illustration of the modeling framework, its components and their relationships. Here SPRL stands for self-play reinforcement learning.}
\label{ModelingFramework}
\end{figure}

The high level conceptual modeling framework and the game playing setting is illustrated by the block diagram in Figure.\ref{ModelingFramework}.
The transportation authority, who decides on the rules of the game, resides in Box 1. All other game participants along with their preferences, which underlie their behaviors in the game, are represented by Box 2. The physical environment, such as roads, initial conditions, communication channels, physical capabilities and constraints, are represented in Box 3. The endogenized behavioral outcomes derived by solving the game with a prescribed algorithm of a chosen solution concept is denoted in Box 4. The conceptual framework illustrated above is very broad. It will enable us to address the following types of questions systematically, depending on the specific perspective of what is treated as input and what is treated as output:
\begin{itemize}
\item The forward problem (given boxes 1, 2, and 3, solve for box 4): solving driving behaviors given the modeling environment and physical constraint, preference of each agent, and the rules of the game. Examples include path planning such as: behaviors in merging, yielding, lane changing, following, passing, and exiting; interactions between different types of agents such as vehicle-pedestrian and vehicle-infrastructure; communication and coordination among agents, and so on.
\item The inverse problem (given boxes 1, 3, and 4, solve for box 2): calibrating decision model parameters given the observed actions from all agents involved. This is critical if the computational framework is going to be quantitatively relevant to reality. One focus here is to understand individual and distributional priorities in various preferences, such as how human drivers rank the relative importance in obeying speed limit, vehicle smoothness, lane deviation, accident avoidance, and how these priorities manifest themselves in terms of the appropriate functional forms.
\item The mechanism design problem (same as the forward problem, but with a feedback loop from box 4 to solve for box 1): refining and optimizing the game rules from the transportation authority's point of view. Topics in this area can include how to tweak and improve existing traffic rules and how to invent the new ones to accommodate new vehicle capabilities and the ever changing new infrastructure; how to assign objective functions for algorithmic agents, such as autonomous vehicles; how to handle potential moral ambiguities which can arise.
\end{itemize}

In this paper we will solely concentrate on the forward problem, and leave the inverse and mechanism design problems to our forthcoming papers \cite{CalibrationPaper} and \cite{CAVandShockwaves}.

\subsection{Modeling Environment}\label{ModelingEnvironment}
There are multi-agent $i\in I=\{1, 2, ..., N\}$ playing the game specified by the transportation authority. Time is discrete: $t\in\{0, 1, 2,...,T-1\}$ in units of $\Delta t$. The precise value of choice for $\Delta t$ is based on a compromise among several factors: 1) making the dynamics sufficiently smooth and safe; 2) making the planning time horizon $T\times\Delta t$ not too small so that there can be some room for anticipation; and 3) $\Delta t$ being reasonably close to human reflex time during driving. Our typical choices are $\Delta t=0.1 \sim 0.2$ second, and $T\times\Delta t$ is typically $\sim 10$ seconds. 

We follow the standard non-cooperative game theory approach: decision-making is modeled as utility maximizing. The state of agent $i$ at period $t$, $s_{i,t}$, embodies sufficient information for decision-making. In our setting, $s_{i,t}$ typically contains  surrounding agents' positions and velocities in the simplest cases. It may include additional information in more complex settings, such as the orientation of a vehicle, and indicators for certain events, e.g., whether a vehicle has made a stop at a red-light. The action for agent $i$ at time $t$, $a_{i,t}$, is typically a vector including its own acceleration at time $t$, steering angle, turning on signaling light, and so on. Again, the details highly depend on the setting. For convenience, we find that it is easier to separate the private information, such as the destination, from the observable state variable. When necessary we will detail the specific communication mechanism for sharing private information.

In a multi-agent setting, it is not so obvious {\it a priori} who should have the strategic precedence, except for the transportation authority. Therefore, the theoretical framework of discrete time Markov sequential game with simultaneous move \cite{Fudenburg96} appears to be the most relevant choice as our starting modeling framework. Generally, the game should have been treated as a stochastic one, which would require solving coupled Markov Decision Problems. Unfortunately, it is well known that even a MDP with a single decision maker is already very hard to be made computationally practical in realistic settings, let alone the coupled one. Therefore, we need to find a way to dramatically reduce the computational burden if our approach has any chance of being useful at all, apart from purely theoretical exercises. To this end, we introduce an explicit risk premium into the utility function, borrowing a concept from economics. The basic idea is to let the agents involved feel the risk, in the sense of decision-making, before any actual physical damage occurs. For example, even when two cars have not yet touched each other, their drivers can feel the danger of collision when the distance between them is sufficiently small. It is this danger that will be explicitly reflected in the utility function.\footnote{Alternatively, we can also view the risk premium introduced here as a reward shaping \cite{RewardShaping}, a well known technique in reinforcement learning literature, for the sparse penalty associated with collisions.} Once the risk premium is introduced, the precise form of uncertainties become less critical. This in turn allows us to treat the game deterministically, thereby radically simplifying the game. We intuitively believe that our modeling strategy is consistent with human driving heuristics.\footnote{We comment on how to reconcile our deterministic formulation with the standard MDP approach in subsection \ref{IncorporatingUncertainties}, where we explicitly introduce uncertainties in the state evolution and allow random deviations from the optimal action.}

Since the game we consider is Markov with a continuous and deterministic state evolution, the solution space is limited to Markov pure strategies. The Markov condition implies that the per period utility function for agent $i$ at time period $t$, with own action $a_{i,t}$ and actions of others $a_{-i,t}\,$, can be written as $u_{i,t}(a_{i,t}|s_{i,t},a_{-i,t})$, i.e., it only depends on the current state and actions. The dependence of agent $i$'s utility on its own action $a_{i,t}$ is obvious. The need for the dependence on others' action $a_{-i,t}$ is for the purpose of handling interactions among agents. Furthermore, we assume that it can be decomposed as a sum over a number of components with appropriate weights\footnote{This decomposition is standard in both economics and in machine learning literature, see for example \cite{MaxEntropyIRL, InverseRL, ImitationLearning}.}
\begin{equation}
u_{i,t}(a_{i,t}|s_{i,t},a_{-i,t})=\sum_k \,w_{i,k}\,\phi_{i,t}^{(k)}(a_{i,t}|s_{i,t},a_{-i,t})\, ,
\end{equation}
where the weights are agent dependent in order to take care of potential heterogeneity, but not time dependent within the planning time horizon. Different problem could imply different set of components to be explicitly included in the sum. Some of the components can represent reward, such as moving towards the goal, whereas others can represent penalties, such as collision, lane violation, roughness. We anticipate that the weights will form a hierarchy in terms of their magnitudes ranging from small to very large, reflecting the preferred priority of the agent. Furthermore, the Markov condition implies that the state evolution can be written as
\begin{equation}
s_{i,t+1}=f_{i,t}(a_{i,t}|s_{i,t},a_{-i,t})\, ,\label{StateEvolution}
\end{equation}
where $f_{i,t}()$ is a deterministic function specified by the appropriate Newtonian kinematics or dynamics. This function can be dependent on agent $i$ and potentially on the explicit time period $t$. 

It is conceptually important to realize that we treat the utility function purely as a decision modeling device. The utility function can deviate from the physical benefits and penalties, especially the part related to risk premium. Consequently, the precise form of the risk premium will depend on the solution concept used to derive the behavioral outcome. For example, the risk premium deployed to solve the exact Nash equilibrium needs not be the same as that used in a method for deriving a heuristic solution to approximate the Nash equilibrium. This is because that, at least for deterministic pure strategy games, everything is perfectly anticipated in a Nash equilibrium, whereas the time horizon of the anticipation in a heuristic solution is typically shorter than the full planning time horizon. Therefore, the utility function should be commensurate with its corresponding solution concept. Of course, the physical part of the utility should always be kept identical. Ultimately, the utility function should always be calibrated using the actual traffic data together with the appropriate solution concept.

In the next section, we first explore the solution concept based sub-game perfect Nash equilibrium. In section \ref{SolutionConcept2}, we will show how to relax the deterministic assumption by explicitly introducing noise factors when we replace the Nash equilibrium based solution concept with a heuristic based concept of adaptive optimization with finite look-ahead anticipation.

\section{\bf Solution Concept I: Sub-Game Perfect Nash Equilibrium } \label{SolutionConcept1}
In this and the following sections we mostly keep things general and formal. Implementation details will be illustrated in Section \ref{SimulationExperiment} when we apply the proposed algorithms to a concrete example.

\subsection{Theoretical Setup}
Here we deploy the standard non-cooperative game theory solution concept in a deterministic dynamic setting: Markov sub-game perfect Nash equilibrium. The aim of each agent is to maximize the cumulative utility function defined as 
\begin{equation}
U_i(a_{i,\{t=0,1,...,T-1\}}|s_{i,0},a_{-i,\{t=0,1,...,T-1\}})=\sum_{t=0}^{T-1}\, u_{i,t}(a_{i,t}|s_{i,t},a_{-i,t})\, ,\label{CumulativeUtility}
\end{equation}
while taking into account how other agents would behave using only Markov strategies. In the above equation we have used the fact that future states, $s_{i,t>0}$, can be expressed as a function of the initial condition $s_{i,0}$ and the action sequences via the state evolution Eq.(\ref{StateEvolution}). Since the planning time horizon in our context is typically a few seconds there is no point to introduce discounting.

The best response of agent $i$ to a given set of fixed action sequence of all other agents, $a_{-i,\{t=0,1,...,T-1\}}$, is then defined by maximizing the cumulative utility function with respect to a sequence of self-action sequence
\begin{equation}
a_{i,\{t=0,1,...,T-1\}}^*(s_{i,0},a_{-i,\{t=0,1,...,T-1\}})=\argmax_{a_{i,\{t=0,1,...,T-1\}}}
U_i(a_{i,\{t=0,1,...,T-1\}}|s_{i,0},a_{-i,\{t=0,1,...,T-1\}})\, . \label{BestResponse}
\end{equation}
Note that the above equation effectively becomes a ``static but simultaneous" optimization problem for all the self-action sequence $a_{i,t}\, ,\,\forall t\in\{0,1,...,T-1\}$, hence side-stepping all the computational issues associated with dynamic programming. In particular, the curse of dimensionality and representation of value function do not arise in our context. The Nash equilibrium of the game is reached when all agents are simultaneously using their best responses against each other, i.e.,
\begin{equation}
a_{i,\{t=0,1,...,T-1\}}^*(s_{i,0},a_{-i,\{t=0,1,...,T-1\}}^*)=\argmax_{a_{i,\{t=0,1,...,T-1\}}}
U_i(a_{i,\{t=0,1,...,T-1\}}|s_{i,0},a_{-i,\{t=0,1,...,T-1\}}^*)\, . \label{NashEquilibrium}
\end{equation}
This condition is particularly useful for verifying whether the solution found by some numerical procedure is actually a Nash equilibrium. 

One subtle aspect of our formulation above is that we only impose the initial condition explicitly in the optimization process, while leaving the terminal condition $s_{i,T}$ implicit, essentially as the outcome of the optimization. We choose to enforce the terminal condition, such as destination, in a soft manner in the sense of Lagrange, by augmenting the utility function. This choice generally makes the optimization problem easier, but also endogenizes certain terminal outcomes, such as avoiding accidents or forming traffic jams when reaching the destination becomes physically impossible.

We can also re-write the optimization problem in Eq.(\ref{BestResponse}) as the following
\begin{equation}
W_{i,t}(s_{i,t}|a_{-i,\{t,...,T-1\}})=\max_{a_{i,t}}\Big\{ u_{i,t}(a_{i,t}|s_{i,t},a_{-i,t})+W_{i,t+1}(s_{i,t+1}=f_{i,t}(a_{i,t}|s_{i,t},a_{-i,t})|a_{-i,\{t+1,...,T-1\}})
\Big\}\, , \label{BellmanEquation}
\end{equation}
with the utility-to-go function $W_{i,t}()$ is defined, in conjunction with the state evolution, as
\begin{equation}
W_{i,t+1}(s_{i,t+1}|a_{-i,\{t+1,...,T-1\}})=\max_{a_{i,\{t+1,...,T-1\}}}\Big\{ \sum_{t'=t+1}^{T-1}
u_{i,t'}(a_{i,t'}|s_{i,t'},a_{-i,t'})\Big\}\, . \label{UtilityToGoFunction}
\end{equation}
Eq.(\ref{BellmanEquation}) is immediately recognized as the Bellman equation from dynamic programming. This in turn implies that the above Nash equilibrium defined in Eq.(\ref{NashEquilibrium}) is also a sub-game perfect equilibrium.

\subsection{Algorithm I: {\it betaNash}}
We exploit the best response dynamics to numerically solve the Markov sub-game perfect Nash equilibrium with pure strategy. The basic idea is that, starting from some reasonable initial action sequence, each agent tries to respond optimally in every iteration (viewed as self-play learning process) given the prevailing strategies of all other agents. Under certain mathematical conditions, the best response dynamics converges to a Nash equilibrium. A broad class of such games, called supermodular games with pure strategy, can be found in economics \cite{Milgrom90}. Formally, the best response dynamics is defined by
\begin{equation}
a_{i,\{t=0,1,...,T-1\},\tau+1}^*(s_{i,0},a_{-i,\{t=0,1,...,T-1\},\tau})=\argmax_{a_{i,\{t=0,1,...,T-1\}}}
U_i(a_{i,\{t=0,1,...,T-1\}}|s_{i,0},a_{-i,\{t=0,1,...,T-1\},\tau})\, , \label{BestResponseTau}
\end{equation}
where we have added subscript, $\tau$, to denote the iteration step for the best response dynamics. It is worth pointing out that, while there was no explicit coordination in the formulation, the actual coordination is achieved from the iterative nature of the best response dynamics from the above equation. In some sense, the best response iteration provides a ``negotiation process'' for all agents involved to figure out one another's intentions, such as where to go in what order at what speed, so that a mutually satisfactory outcome is obtained for all agents involved.

We also recognize that the best response dynamics is a specific form of self-play reinforcement learning originally invoked to justify why Nash equilibrium, which is so complex mathematically, exists and can be found by its game players that have generally bounded rationality. It is also very similar, at least in spirit, to the self-play reinforcement learning used in DeepMind's alphaGo Zero \cite{alphaGo}, though the detailed state representation, function approximation, and search strategies are totally different. There is no need for a neural network representation in our context, due to the fact that we know the explicit functional forms for utility functions and state evolution. It is much more efficient computationally to rely on traditional maximization methods than on stochastic gradient based or tree-search methods in the best response dynamics.

With the above mathematical preparation we state the algorithm below:

\IncMargin{1em}
\begin{algorithm}[H] 
	\DontPrintSemicolon
	\SetKw{True}{true}\SetKw{Break}{break}
	initialize the simulation environment\;
	\For{$\forall i\in I$}{
		initialize state variables $s_{i, t = 0}$\;
		initialize the action set, e.g. $a_{i, \{t = 0, 1, \cdots, T-1\}, \tau=0} = 0$\;		
	}
	\For(\tcp*[h]{this is the loop for best response dynamics}){iteration step $\tau = 1: n$}{
		\For{agent $ i = 1: N $}{
			get the most recent action sequence:  $a_{-i,\{t=0,1,...,T-1\},\tau-1}$\;
			do the optimization for agent $i$: \tcp{the loop over $T$ is embedded in this step}
			\nonl\quad$a^{*}_{i,\{t=0,1,...,T-1\},\tau} =\argmax_{a_{i,\{t=0,1,...,T-1\}}}U_i(a_{i,\{t=0,1,...,T-1\}}|s_{i,0},a_{-i,\{t=0,1,...,T-1\},\tau-1})$\;
			update the action set:  $a_{i,\{t=0,1,...,T-1\},\tau} \leftarrow a^{*}_{i,\{t=0,1,...,T-1\},\tau}$ \;
		}
		check Nash equilibrium condition:
		$
		a_{i,\{t=0,1,...,T-1\}}^*(a_{-i,\{t=0,1,...,T-1\}}^*)=\argmax_{a_{i,\{t=0,1,...,T-1\}}}
		U_i(a_{i,\{t=0,1,...,T-1\}}|a_{-i,\{t=0,1,...,T-1\}}^*)
		$\;
		\If{condition is \True}{\Break}
	}
	\For{$\forall i\in I$}{
		\For{$ t = {0, 1, \cdots, T-1}$}{
			$s_{i, t+1} = f_{i, t}(a^*_{i, t}| s_{i, t}, a^*_{-i, t})$
		}
	}
	\caption{\it betaNash}\label{betaNash}
\end{algorithm}\DecMargin{1em}

\vspace{4 mm}
\noindent The algorithm defined above ({\it betaNash}) has a number of nice properties:
\begin{itemize}
\item It is based on a rigorous Markov game whose Nash equilibrium can be solved by a systematic numerical method called best response dynamics.
\item The dynamic Nash equilibrium so obtained is automatically sub-game perfect by construction.
\item The best response dynamics provides an explicit negotiation or learning mechanism among all agents, for figuring out both their driving intentions and optimal actions.
\item While the assumptions for game theory is very strong so that the derived sub-game perfect Nash equilibrium solutions may not be immediately relevant for modeling mixed settings where human drivers are involved, they at least serve as the theoretical benchmark. We intuitively expect that solutions based on game theory are often more efficient in coordination than heuristically motivated solutions. This is also what we experienced empirically. One such example will be presented in Section \ref{SimulationExperiment}. Other examples can be found in our forthcoming papers \cite{CoordinationPaper, AutoValetParking}.
\item It is also conceivable, in the forthcoming connected and autonomous world, that the same solutions can be implemented through an edge infrastructure system, provided that the decision-making for all vehicles in the vicinity can be centralized. 
\item The algorithm is computationally feasible for at least some small but realistic problems where the number of agents in the game is not too large.
\end{itemize}
However, the algorithm also has a number of weaknesses:
\begin{itemize}
\item The assumptions associated with the dynamic Nash equilibrium may be too strong to be sufficiently realistic. Of course, this is the same criticism for the relevance of the basic assumptions for game theory, such as the common knowledge (e.g. knowing the utility functions for all agents) and complete rationality (e.g. infinite computing power).
\item The common knowledge assumption perhaps can only be realizable when all agents are autonomous, whose utility functions are endowed, and the solution is being derived by a central controller.
\item The computational burden grows dramatically when the number of agents becomes big, at least $N^2\times T$ times the number of iterations in the best response dynamics, where $N$ is the number of agents involved in the game. This makes it hard to be applied to realistic settings where $N$ is large.
\item The best response dynamics is intrinsically sequential as defined in Eq.(\ref{BestResponseTau}), and hence {\it betaNash} is not naturally parallelizable. It is possible to make the algorithm parallel by simultaneously calculating the best response to the actions of all other agents at one time step earlier, though this may lose some efficiency on a per iteration basis.
\item Another issue, though conceptually minor, is the number of agents $N$ has to be fixed for the game in the entire planning time window. This is not very realistic, given the fact the planning horizon can be as long as $\sim 10$ seconds.
\item The common knowledge and rationality assumptions, together with the heavy computational burden, will also limit the scope of applications in the inverse problems and mechanism design problems.
\end{itemize}

It is in hoping to overcome these weaknesses that prompted us to pursue another solution concept, as detailed in the next section.


\section{\bf Solution Concept II: Adaptive Optimization with Finite Look-Ahead Anticipation} \label{SolutionConcept2}
In the preceding section, we introduced an explicit risk premium into the utility function so that the would-be stochastic game can be simplified to a game of perfect information, at least from the point of view of the state evolution. Consequently, the computational burden is eased substantially. However, we have additional reasons to further simplify the model so that we can accommodate human drivers and autonomous vehicles at the same time.  This is due to several considerations. First, once in a mixed setting, where human agents and algorithmic agents coexist, the common knowledge assumption becomes dubious. Of course, we can always revoke the Bayesian Nash equilibrium concept to relax some of the common knowledge requirements. But that will again substantiate the computational burden. Second, it is unreasonable to assume that human agents, with bounded rationality, can always find the pertinent Nash equilibrium in real-time in realistic traffic settings. Third, in order to make the approach practical, further reduction in computational requirement is necessary.

Before we proceed more formally let us first recapitulate human driving heuristics:
\begin{itemize}
\item All decisions are decentralized and individually made by each agent, mostly without explicit communication.
\item All the immediate actions are planned according to the current state, with some reasonable or good enough but otherwise imprecise anticipation for the future state.
\item In case the future state has multiple scenarios, action planning is done according to the most conservative one.
\item The planned actions have some time persistence, unless they are interrupted by environment change.
\item The action planning starts anew once the new information is assessed.
\end{itemize}

\subsection{Theoretical Setup}
Armed with the formulation in section \ref{SolutionConcept1}, we have a good starting point from which we can gradually relax the conditions that were originally required for the Nash equilibrium approach. We continue to rely on human driving heuristics to guide us in this endeavor. The focus is to shift from the perspective of the full game and the pertinent equilibrium in the preceding section to the perspective of individual agent's decision-making in this section.

The first step is to simplify the state evolution in Eq.(\ref{StateEvolution}). Recall that we used the state evolution to eliminate all the dependence of the cumulative utility function on future states $s_{i,t>0}$. While this is natural theoretically, it is hard to imagine that a human driver can iterate the state evolution many times in a realistic traffic setting, in which both one's own and others' future action sequences are needed. More likely, a human driver will simply anticipate the future states by extrapolating from the current state and the immediate own action and some reasonable assumptions on action sequences for all other agents subsequently. This motivates the following estimated state evolution
\begin{equation}
\tilde{s}_{i,t+1}=\tilde{f}_{i,t}(a_{i,t}|s_{i,t})\, . \label{StateEvolutionTilde}
\end{equation}
In the above equation we have kept the self-action dependence and dropped the precise dependence on actions for all other agents by assuming $a_{-i,t>0}$ being some naturally intuitive maneuvering that is commensurate with the specific context. However, Eq.(\ref{StateEvolutionTilde}) cannot be entirely independent of $a_{-i,t>0}$, otherwise we would have been dealing with non-interacting cases. We specify this imprecise dependence on $a_{-i,t>0}$ in Eq.(\ref{StateEvolutionTilde}) by what we call anticipation for agent $i$ at time $t$ given state $s_{i,t}$. Our strategy is to replace the dynamically interacting sequence of actions precisely by some prescribed sequence of actions for all the other agents that are reasonably consistent with the context. In the remainder of this subsection we outline the two generic elements necessary for finite look-ahead anticipation. Of course, the detailed form for the anticipated $a_{-i,t>0}$ is highly context dependent and hence cannot be made general. We will provide a concrete example on how this is done in the simulation experiment in Section \ref{SimulationExperiment}, including algorithmic details.

The next step is to simplify the best response in Eq.(\ref{BestResponse}). One obvious approach is to break the planning horizon into smaller chunks. The extreme case is to let $T=1$. Unfortunately, we know empirically that purely myopic decision-making is not sufficient. Therefore, some form of anticipation needs to be built into the process. Towards this end, we define an effective utility function by looking ahead $h\gg 1$ periods: 
\begin{equation}
\tilde{u}_{i,t}(a_{i,t}|s_{i,t};h)=\sum_{k}\, w_{i,k} \,g_k\Big(\phi_{i,t}^{(k)}(a_{i,t}|\tilde{s}_{i,t});h\Big)\, .\label{EffectiveUtilityTilde}
\end{equation}
The choice of functional form for $g_k()$ depends on the specific feature and is made conservatively when necessary for safety reasons. For some components such as the moving forward reward and lane departure penalty, $g_k()$ are the average of the corresponding components in $h$-period. For driving smoothness, we choose the penalty in the first period. For components that are potentially calamitous, such as crash or collision penalties, we choose maximum penalty among the $h$ periods given the prescribed action sequences for others. The effective utility function defined in Eq.(\ref{EffectiveUtilityTilde}) embodies the first element of what we call finite look-ahead anticipation once all the paths are hypothesized. 

Then, the resulting approximate best action can be derived from 
\begin{equation}
\tilde{a}_{i,t}^*(s_{i,t};h)=\argmax_{a_{i,t}} \tilde{u}_{i,t}(a_{i,t}|s_{i,t};h)\, . \label{BestResponseTilde}
\end{equation}
Notice that, due to the dropping of the action sequence for all other agents in the simplified effective utility function, the above optimization is only a reaction to the pre-chosen action sequences in the anticipation, and hence the original game problem is reduced to a control problem step by step. 

One subtlety arises when there is no explicit coordination mechanism in Eq.(\ref{BestResponseTilde}), how could each agent figure out the intentions of all the other nearby agents in case they do not have the common knowledge? To address this problem, we introduce the concept of path scenario: a finite number of possible paths that a nearby agent can pursue at any give moment. This can be done by enlarging the state space in Eq.(\ref{BestResponseTilde}). Again, we take a conservative approach to plan the action for agent $i$ with the most conservative path scenarios from all nearby agents. In this way, the state variable should be understood also as adaptive, including the number of nearby agents and their path scenarios. This can be viewed as the second element of finite look-ahead anticipation. It is important to emphasize that the anticipation of future state is done for all agents, including self and all other agents, though they are treated differently, due to the fact that the self intention is known whereas others' intentions are unknown.

Once solved from Eq.(\ref{BestResponseTilde}) the solution $\tilde{a}_{i,t}^*(s_{i,t};h)$ is only executed for one period, even though it is derived with information from $h$ periods. As time increments from $t$ to $t+1$, a new optimization problem is re-assessed from the observable state variables at $t+1$. This is our third step in formulating a modeling framework which is ultimately capable of handling human driving behaviors and being computationally feasible at the same time. The intuition behind this modeling strategy is that each agent plans its path at micro level with deliberate intention and anticipation, and then monitors the situation constantly and adapts when necessary. Of course, questions such as whether such a modeling strategy can truly approximate human driving behaviors well and whether it is sufficient to avoid accidents, can only be answered empirically. Our initial applications using solution concepts outlined in this section in a variety of settings \cite{CalibrationPaper, CAVandShockwaves, CoordinationPaper, AutoValetParking, PedestrianPaper, RoundaboutCalibration} indicate that answers are affirmative, thanks to the explicit introduction of the risk premium to the utility function, finite look-ahead anticipation, and the adaptivity. 

Finally, while it is important to demand $h\gg 1$, we should not make $h$ too large. This is because the decision will be dominated by those future time periods where the assumption of $a_{-i,t>0}=0$ becomes totally dubious when $h$ is extremely large. Generally, $h$ should be treated as a hyper-parameter of the model and tuned in each specific modeling situation. However, our experience indicates that $h\times\Delta t$ being in the range of $1$ to $3$ seconds is a good choice, providing robustness for many different driving settings, depending on the specific traffic condition, such as local or highway, crowded or uncrowded, good or bad weather, and so on. We also found empirical evidence that $h$ should be in this range when we calibrate the utility function using traffic video data \cite{CalibrationPaper}.

\subsection{Algorithm II: {\it adaptiveSeek}}
We are now ready to state the algorithm that implements the theoretical considerations in the preceding subsection.

\IncMargin{1em}
\begin{algorithm}[H]
	\DontPrintSemicolon
	\SetKw{True}{true}\SetKw{Break}{break}
	\For{$\forall i\in I$}{initialize the simulation environment}
	\For{$t=0:T-1$}{
		\For(\tcp*[h]{this is the loop over all agents}){$\forall i\in I$}{
			get current state $s_{i, t}$\;
			\For(\tcp*[h]{this is the loop over agents other than $i$}){$\forall j\in I_{-i} $}{
				\uIf{certain desired paths are assessed}{
					anticipate possible path scenarios that agent $j$ likely to take based on its state\;
					\For{$\tilde{t}=0:h-1$}{
						extrapolate trajectory using a prescribed action consistent with the path scenario\;
					}
				}
				\Else{
					assume agent $j$ keep the natural extension of its current state\;
					\For{$\tilde{t}=0:h-1$}{
						extrapolate trajectory using zero action\;
					}
				}
			}
			calculate the effective utility given all the anticipated path scenarios: $\tilde{u}_{i,t}(a_{i,t}|s_{i,t};h)=\sum_{k}\, w_{i,k} \,g_k\Big(\phi_{i,t}^{(k)}(a_{i,t}|\tilde{s}_{i,t});h\Big) $\;
			choose the effective utility with the safest path among all the anticipated path scenarios\;
			do the optimization	$\tilde{a}_{i,t}^*(s_{i,t};h)=\argmax_{a_{i,t}} \tilde{u}_{i,t}(a_{i,t}|s_{i,t};h)\, . \label{BestResponseTilde1}$\;
			\If{add noise}{
				$\hat{a}_{i,t}(s_{i,t})=\tilde{a}_{i,t}^*(s_{i,t};h)+\epsilon_{i,t} $\;
			}
		}
		\For{$\forall i\in I$}{
			apply the observed action and evolve to new states:
			$s_{i, t+1} = f_{i, t}(\hat{a}_{i, t} | s_{i, t}) $\;
		}
	}
	\caption{\it adaptiveSeek}\label{adaptiveSeek}
\end{algorithm}\DecMargin{1em}

\vspace{4 mm}
\noindent A few general comments are in order for the above algorithm ({\it adpativeSeek}):

\begin{itemize}
\item Since the decision-making for each agent is independent without explicit coordination in {\it adaptiveSeek}, there is no need to assume that all agents in the game follow the same algorithm. It could be that some agents are human driven, some are autonomous with their own built-in algorithms, so long as all the implied behaviors can be reasonably anticipated. Therefore, we expect that {\it adaptiveSeek} can be used to model mixed settings where all types of agent co-exist.
\item Due to the lack of common knowledge we cannot assume that each agent knows the intentions of all other agents. Consequently, this requires some form of anticipation for the intentions of the agents in the vicinity. We choose to accomplish this with the simplest possible approach - using a prescribed action sequence consistent with the path scenario. Then, the algorithm picks the safest optimal action to execute among all the possible scenarios. In this way, we are able to handle the mixed environment. The {\it if}-{\it else} condition in the optimization process is a direct consequence of the fact that each agent does not have information about others' intentions. 
\item However, due to the inclusion of the anticipation for all the nearby agents in the effective utility function, {\it adaptiveSeek} is still implicitly  coordinated, in contrast with {\it betaNash} where the coordination is explicit. Without some form of coordination, it is not possible to take care of the interactions among all agents.
\item With this fully adaptive approach, we can easily handle any agents entering or leaving the simulation environment at any given time, which makes it more flexible than the solution concept in Algorithm {\it betaNash}.
\item By construction, this algorithm is local, because the utility function depends only on the nearby agents, and hence is linear in the number of agents involved in the game\footnote{It is seemingly an order $N^1$ operation in the loop over $I_{-i}$, which would in turn make {\it adaptiveSeek} an order $N^2$ algorithm. Fortunately, we can make the loop over $I_{-i}$ an order $N^0$ operation if we enlarge the state space by keeping track of a list of all neighboring agents and possibly their neighbors.}. Furthermore, the inner loop can be easily parallelized, as the decision-making for each agent is practically independent from one another at each time step, If the number of processors is big enough, the algorithm is practically order $N^0$. Alternatively, this inner-loop optimization can be distributed to each agent's own on-board computer. It is these nice properties that make {\it adaptiveSeek} very fast. This in turn allows a good chance for it to be made real-time in realistic applications.
\end{itemize}

\subsection{Incorporating Uncertainties}\label{IncorporatingUncertainties}
So far we have been solely concentrating on the deterministic limit. Now we would like to re-introduce uncertainties explicitly to our formulation. There are several reasons that motivate us for this: 1) checking robustness of the deterministic modeling results against small disturbances; 2) solving the inverse problems using techniques developed in inverse reinforcement learning with maximum likelihood \cite{InverseRL, ImitationLearning} or maximum entropy methods \cite{MaxEntropyIRL}; and 3) designing good mechanisms for improving traffic rules or regulations. To that end, we need to distinguish the perceived state by the decision-maker from the actually realized state. We denote the former by $\hat{s}_{i,t}$ and continue to use $s_{i,t}$ to denote the latter. Likewise, we also need to distinguish the actually realized action, $\hat{a}_{i,t}$, from the intended optimal action $a^*_{i,t}$. Note that the intended optimal action is now a function of the perceived state, not the actual state, i.e. $a^*_{i,t}=a^*_{i,t}(\hat{s}_{i,t})$. With these new notations, we can re-state the equations related to the action optimization and state evolution for Algorithm {\it adaptiveSeek} as follows (c.f. Eq.(\ref{StateEvolutionTilde}, \ref{BestResponseTilde})):
\begin{equation}
\begin{cases}
\hat{s}_{i,t}&=s_{i,t}+\epsilon^s_{i,t}\, ,\\
a_{i,t}^*(\hat{s}_{i,t};h) &=\argmax_{a_{i,t}} \tilde{u}_{i,t}(a_{i,t},\hat{s}_{i,t};h)\, , \\
\hat{a}_{i,t}(\hat{s}_{i,t}) &=a_{i,t}^*(\hat{s}_{i,t};h)+\epsilon^a_{i,t}\, , \label{BestResponseObs} \\
s_{i,t+1}&=\tilde{f}_{i,t}(\hat{a}_{i,t}(\hat{s}_{i,t})|s_{i,t})\, ,
\end{cases}
\end{equation}
with $\epsilon^s_{i,t}$ and $\epsilon^a_{i,t}$ obeying some chosen statistical distributions. While we included the noise terms additively in the above equation for notational simplicity, it might be more appropriate to treat some of them multiplicatively. Notice that the realized action is inflicted by two sources of uncertainty, one explicitly in $\epsilon^a_{i,t}$, the other implicitly through the perceived state $\hat{s}_{i,t}$ in $a^*_{i,t}(\hat{s}_{i,t})$. Since the state evolution is supposed to be intrinsic, not from the decision-making perspective, the actual state evolves from the actual initial state $s_{i,t}$ under the realized action $\hat{a}_{i,t}(\hat{s}_{i,t})$. The origins of the noise terms in Eq.(\ref{BestResponseObs}) can be either due to instrument noise, human estimation and execution errors from the idealized state and optimal action or due to minor uncertainties of the road, such as local slopes, bumps and potholes, poor weather and lighting conditions. The specific distributional assumption depends on the context of the modeling. These can be as simple as IID normal distributions, or they can be more complex, such as containing some form of auto-regressive structure. 

In some sense, the above specification reconciles our earlier deterministic formulation posteriorly with the common MDP formulation for the game, where the state evolution is stochastic and action can take on mixed strategy. Viewed through a similar lens, we can also regard the deterministic formulation in subsection \ref{ModelingEnvironment} as an approximation to the MDP specification using certainty equivalent approach that was well known in adaptive control literature \cite{BertsekasBook}. The certainty equivalent approximation aims to substantially reduce the computational burden by replacing probabilistic expectations with their corresponding expected values, in the spirit of $E[V(\cdot,\epsilon)]\approx V(\cdot,E[\epsilon])$ or its variations, so that the policy optimization becomes deterministic, as what we did in subsection \ref{ModelingEnvironment}.

Similarly, though with additional notational complication, we can also re-write the state evolution and action optimization for Algorithm betaNash in section \ref{SolutionConcept1}. In this case the sub-game perfect Nash equilibrium conditions need to be understood as that under the certainty equivalent approximation mentioned above. Even so, the equilibrium has to be solved period by period, not only to be adaptive to the realized state, which is now contaminated by noises or errors, but also to accommodate the potential change of agents involved in the game at any given moment. 

One can immediately recoganize that Eq.(\ref{BestResponseObs}), along with the kinematic bicycle model in Eq.(\ref{StateEvolutionBicycleModel},\ref{HeadingAngle},\ref{SpeedEvolution}) to be presented in the next section, is in the form of state space models (SSM). The inverse problem, calibrating the utility function using observed action data, is well formulated and investigated in SSM, see for example \cite{TimeSeriesBook}. The driving decisions made by the driver can be incorporated naturally as the control input in SSM. In contrast with the standard inverse reinforcement learning techniques, such as those in \cite{Schwarting19, MaxEntropyIRL, InverseRL,ImitationLearning}, where coupled dynamic programming problems have to be solved, the inference using SSM with {\it adaptiveSeek} as the behavioral model only requires solving decoupled static optimization problems. The dramatic computational simplification is possible, because {\it adaptiveSeek} is only contingent on the observable state by individual agent step by step. The dynamic aspect is implicitly encoded in the sequence of the state for each agent, while inter-agent interactions also reside in the observable state. In \cite{CalibrationPaper} we demonstrate how this is done in explicitly analyzing data from Sugiyama experiment \cite{Sugiyama08}.

\section{\bf Simulation Experiment} \label{SimulationExperiment}

So far our discussion has been quite general and formal. Now we apply these concepts and algorithms to a concrete example in this section. Our aims are 1) to illustrate how the concepts and algorithms are actually used in solving specific problems; 2) to show detailed examples of specification of utility functions; 3) to contrast the solutions derived from two very different algorithms: {\it betaNash} and {\it adaptiveSeek}.

\subsection{Double-Lane Highway with an Unexpected Barrier}
The specific setup of the simulation experiment is partially inspired by the work in \cite{Zhang18}.
\subsubsection{Simulation Environment}
\begin{itemize}
\item Total planning simulation time span is 8 seconds, with decision time interval $\Delta t = 0.2$ second, resulting in $T=40$. 
\item Two lanes in parallel running from left to right, with an unexpected barrier located at $x=0$ blocking the lower lane starting at $t=0$.
\item All lengths are in meters, all velocities are in $m/s$, all accelerations are in $m/s^2$, and angles are in degrees.
\item There are two vehicles running along the highway with constant speed, one in each lane initially.
\item The state is understood broadly that includes all causal information necessary for decision-making at any given moment. These include all the positions, velocities, accelerations, steerings, and orientations of the two vehicles in the immediate past.
\end{itemize}

\subsubsection{State Evolution: Kinematic Bicycle Model}
We adopt the kinematic bicycle model \cite{Rajamani06} with only front wheel steering capability to specify the state evolution. In this model the state variable of vehicle $i$ is characterized by its center of gravity coordinate $(x_{i,t},\,y_{i,t})$ and the orientation (aka heading angle or yaw angle) $\psi_{i,t}$ in a fixed coordinate system. The equation of the motion for vehicle $i$ is explicitly given by
\begin{eqnarray} 
x_{i,t+1}&=& x_{i,t}+\Delta t\,v_{i,t}\,\cos(\psi_{i,t}+\beta_{i,t})\, ;\nonumber \\
y_{i,t+1}&=& y_{i,t}+\Delta t\,v_{i.t}\,\sin(\psi_{i,t}+\beta_{i,t})\, ;\label{StateEvolutionBicycleModel}\\
\psi_{i,t+1} &=&\psi_{i,t}+\Delta t\,\frac{v_{i,t}}{L}\,\cos\beta_{i,t}\,\tan\delta_{i,t}\, ; \nonumber
\end{eqnarray}
where $L=2.88\,m$ is the wheelbase, and $b$ (chosen to be $L/2$ in this simulation) is the distance from the center of mass to the rear axle. $\beta_{i,t}$ is the slip angle (the angle between the velocity of the center of gravity $v_{i,t}$ and the longitudinal axis of the vehicle) defined as
\begin{equation}
\beta_{i,t}=\tan^{-1}\Big(\frac{b}{L}\,\tan\delta_{i,t}\Big)\, ,\label{HeadingAngle}
\end{equation}
where $\delta_{i,t}$ is the steering angle, a decision variable for vehicle $i$. The other decision variable is the acceleration of the center of gravity of the vehicle $\alpha_{i,t}$, which is acted on the velocity evolution directly as
\begin{equation}
v_{i,t+1}=v_{i,t}+\Delta t\, \alpha_{i,t}\, .\label{SpeedEvolution}
\end{equation}

\subsubsection{Specification of the Utility Function}\label{UtilityTwoLaneHighway}
There are eight components for the utility function, whose interpretations and functional forms are described below:
\begin{itemize}
\item A moving forward reward that peaks at the speed limit of the highway $v_0=31\, m/s$. 
$$\phi_{i,t}^{(1)}=1-\Big(\frac{v_{i,t}-v_0}{v_0}\Big)^2\, .$$
\item Three roughness penalties, two for encouraging acceleration/steering smoothness over time, and one for discouraging hard acceleration and braking. 
\begin{eqnarray*}
&&\phi_{i,t}^{(2)}=\big(\alpha_{i,t}-\alpha_{i,t-1}\big)^2\, ; \nonumber \\
&&\phi_{i,t}^{(3)}=\big(\delta_{i,t}-\delta_{i,t-1}\big)^2\, ; \nonumber \\
&&\phi_{i,t}^{(4)}
=\ln\Big(1 + \exp\big[\kappa^{(4)}(\alpha_{i,t}-\bar{\alpha})\big]\Big)
+\ln\Big(1 + \exp\big[-\kappa^{(4)}(\alpha_{i,t}-\underline{\alpha})\big]\Big) \, ; \nonumber
\end{eqnarray*}
where $\bar{\alpha}=4\, m/s^2$ and $\underline{\alpha}=-5\, m/s^2$ are the acceleration and braking limits respectively. The parameter $\kappa^{(4)}=15.0$ represents the hardness of the vehicle's physical constraint on hard acceleration and braking.\footnote{There should also be a similar component for penalizing hard steering. However, due to the range of steering in our simulation experiment being fairly small, we ignore such penalty for simplicity.}
\item A lane departure penalty that incentivizes the vehicle to stay in the middle of either lanes centered at $\pm 1.85\,m$. 
$$\phi_{i,t}^{(5)}=\min\Big[\big(y_{i,t}^2 - (W/2)^2\big)^2/(3 W^4/4), ~1\Big]\, ;$$
where $W=3.7\, m$ is the lane width, and the two lanes are centered at $\pm W/2$.
\item An out of road penalty that gives a huge penalty when the vehicle is off the road shoulder. 
$$\phi_{i,t}^{(6)}=S\Big(\kappa^{(6)}\big(|y_{i,t}|- (W + w/2)\big)\Big)\, ; $$
where $w=2.0\, m$ is the vehicle width, $S(x) = 1/[1 + \exp(-x)]$ the {\it sigmoid function}, and the parameter $\kappa^{(6)}=3.0$ controls how fast the penalty sets in.
\item A crash penalty that at the location of the unexpected barrier $x=0$ at the lower lane. 
$$\phi_{i,t}^{(7)}=S\Big(\kappa^{(7)}_x(x_{i,t} + l_x^{(7)})\Big)\cdot S\Big(-\kappa^{(7)}_y(y_{i,t} - l_y^{(7)})\Big)\, ; $$
where the crash risk premium related parameters $l_x^{(7)}=5.0\,m$ and $l_y^{(7)}=1.0\,m$, and the parameters $\kappa^{(7)}_x=2.0$ and $\kappa^{(7)}_y=20.0$ control how fast the risk premium drop to zero longitudinally and laterally.
\item A collision penalty between (pairwise) nearby vehicles whose shapes are assumed to be rectangles. 
$$\phi_{ij,t}^{(8)}=\Big[ \tilde{S}\Big(\kappa^{(8)}_x(\Delta x_{ij,t} + l_x^{(8)})\Big) + \tilde{S}\Big(\kappa^{(8)}_x(l_x^{(8)}-\Delta x_{ij,t})\Big) \Big] \cdot \Big[\tilde{S}\Big(\kappa^{(8)}_y(\Delta y_{ij,t} + l_y^{(8)})\Big) + \tilde{S}\Big(\kappa^{(8)}_y(l_y^{(8)}-\Delta y_{ij,t})\Big)\Big]\, ; $$
where $\Delta x_{ij,t}=x_{i,t}-x_{j,t}$, $\Delta y_{ij,t}=y_{i,t}-y_{j,t}$, and $\tilde{S}(x)=S(x)-1/2$. The collision risk premium related parameters are
$l_x^{(8)}=10.0\,m$ and $l_y^{(8)}=2.0\,m$, with $\kappa^{(8)}_x=0.5$ and $\kappa^{(8)}_y=9.0$ controlling their longitudinal and lateral ranges.
\end{itemize}
The corresponding weights are: $w_1 = 1.0$ by convention, $w_2 = -0.01$, $w_3 = -1.5$, $w_4 = -1.0$, $w_5 = -0.3$, $w_6 = -24.0$, $w_7 = -20.0$, and $w_8 = -14.0$. For simplicity, all these values are shared by both vehicles in this simulation experiment. Note that the first six components are purely physical self-effects, and only the last two involve interactions and risk premiums.

Admittedly, the above functional forms and parameters in these components and their corresponding weights are intuitively chosen. They are also somewhat arbitrary within a range. Our simulation results are quite robust so long as the parameters do not deviate too much. Ultimately, these functional forms and parameters and weights are supposed to be calibrated using real data of the vehicles and observed driving behaviors. There are certainly other complications. For example, some of the parameters may be dependent on many other factors, such as vehicle speed, weather condition, lighting, and so on. Generally, how best to assess the utility function and possible subtleties which may arise are topics in their own right and hence is beyond the scope of this paper. In two separate papers, \cite{CalibrationPaper, RoundaboutCalibration}, we attempt to start addressing these types of issues more systematically.

\subsection{Solution from {\it betaNash}}
\subsubsection{Initial Conditions}
We derive the sub-game perfect Nash equilibrium solution by using algorithm {\it betaNash} to iteratively solve Eq.(\ref{BestResponseTau}) under the following two initial conditions, designed to obtain qualitatively different merging behaviors:
\begin{eqnarray*}
\text{IC1}: && \Big\{
\begin{array} {rl}
\text{vehicle in open lane:} &\, (x_{1,0},y_{1,0})=(-90,\,1.85)\, ; \,\,\,(v_{1x,0},v_{1y,0})=(31,\,0)\, ;\\
\text{vehicle in blocked lane:} &\, (x_{2,0},y_{2,0})=(-80,\,-1.85)\, ; \,\,\,(v_{2x,0},v_{2y,0})=(31,\,0)\, .
\end{array} \\
\text{IC2}: && \Big\{
\begin{array} {rl}
\text{vehicle in open lane:} &\, (x_{1,0},y_{1,0})=(-80,\,1.85)\, ; \,\,\,(v_{1x,0},v_{1y,0})=(31,\,0)\, ; \\
\text{vehicle in blocked lane:} &\, (x_{2,0},y_{2,0})=(-80,\,-1.85)\, ; \,\,\,(v_{2x,0},v_{2y,0})=(31,\,0)\, . 
\end{array}
\end{eqnarray*}
We further assume zero initial actions: $\alpha_{i,t=-1}=0$ and $\delta_{i,t=-1}=0$. Because we are using {\it betaNash}, which assumes common knowledge, we have to regard the two vehicles being autonomous and controlled by the edge infrastructure. There are 80 decision variables for each vehicle $(\alpha_{i,t},\delta_{i,t})$ with $t\in \{0,\,1,\,\cdots,\,T-1\}$. Intuitively, when the vehicle in the blocked lane leads the vehicle in the open lane in IC1, a front merge is expected to be the more natural outcome. When the vehicle in the blocked lane is no longer leading in IC2, a rear merge is expected to be the more natural outcome. We deliberately leave enough time and space for both vehicles to maneuver so that non-calamitous solutions are feasible.

\subsubsection{The Sub-Game Prefect Nash Equilibrium}
\begin{figure}[!h]
\centering
\includegraphics[width=.90\textwidth]{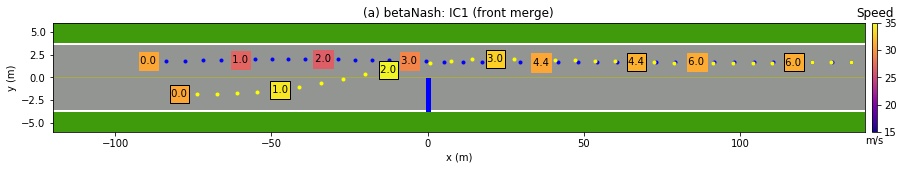} 
\includegraphics[width=.90\textwidth]{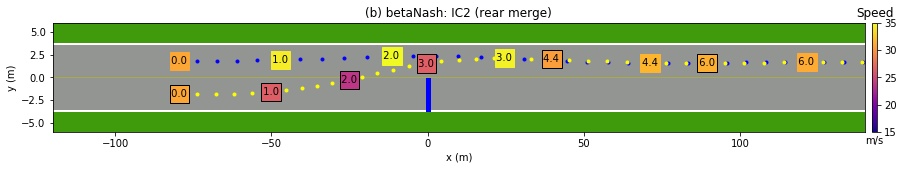} 
\caption{Vehicle trajectories $(x_{i,t}^*,y_{i,t}^*), \forall t\in\{0,\, 1,\cdots 39\}$ in different merging behaviors under different initial conditions in {\it betaNash}: (a) Under IC1, the vehicle in the blocked lane (yellow dot or square with a black frame) first accelerates longitudinally and then turns to the open lane, while the vehicle in the open lane (blue dot or square without a black frame) first yields by slowing down and then accelerates to catch up. (b) Under IC2, the vehicle in the open lane first accelerates and moves to the left slightly, while the vehicle in the blocked lane slows down first, and then turns to the open lane and accelerates to catch up. The number in the boxes represents the time in seconds, and the color represents the speed.}
\label{lane_change_betaNash}
\end{figure}

To avoid local maximum in the best response optimization in Eq.(\ref{BestResponseTau}), we use the {\it Basin-Hopping} algorithm implemented in Python package SciPy, which has some capability for handling global optimization. This appears to be sufficient for {\it betaNash} in our particular setting to converge, thanks likely to the negotiation/learning mechanism inherently embedded in the best response dynamics. The numerical solutions are depicted in Fig.\ref{lane_change_betaNash}, with the upper panel for IC1 and lower panel for IC2. The intuitively expected merging behaviors were borne out explicitly. Indeed, a front merge is the endogenized outcome under IC1, whereas a rear merge is the endogenized outcome under IC2. Note that there was no pre-injected notion of merging behaviors before the solutions are derived. Everything is purely driven by the utility functions and the initial condition. The coordination of the two vehicles in the solutions is entirely achieved during the iterative process of the best response dynamics. The resulting vehicle trajectories and velocities are all smooth. 

\begin{figure}[!h]
\centering
\includegraphics[width=.85\textwidth]{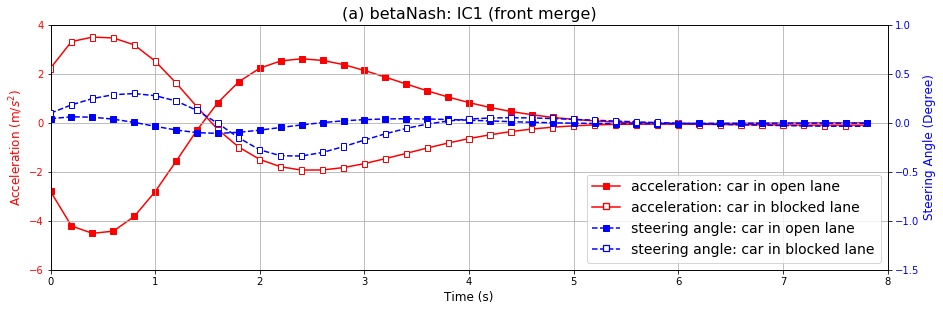} 
\includegraphics[width=.85\textwidth]{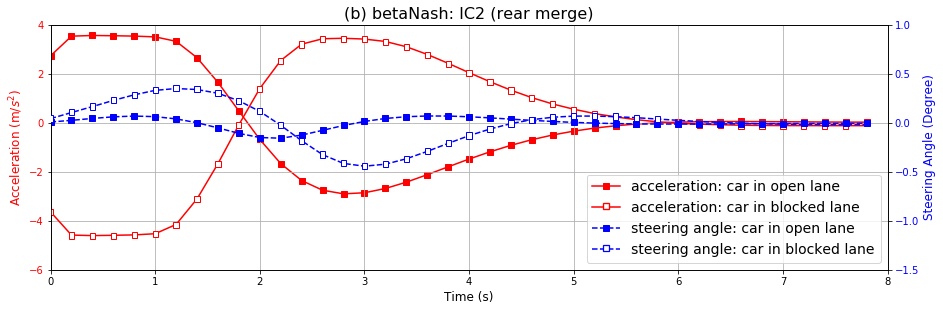} 
\caption{Action sequences for both vehicles in the equilibrium solved by {\it betaNash} under IC1 (front merge) and IC2 (rear merge).}
\label{action_sequence_betaNash}
\end{figure}

In Fig.\ref{action_sequence_betaNash} we plot the action sequences for both vehicles in the equilibrium under IC1 (front merge) and IC2 (rear merge). The solved actions are also smooth in both cases. It is interesting to observe that the vehicle in the open lane swerved to create a little more space for the vehicle in the blocked lane to merge, more so in the case of rear merge than that of front merge.

\begin{figure}[!h]
\centering
\includegraphics[width=0.85\textwidth]{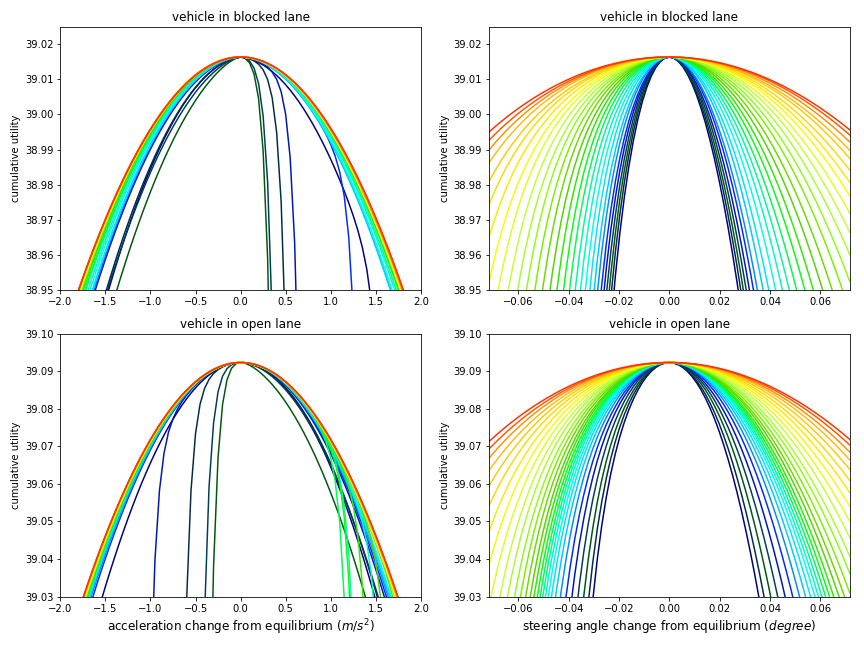}
\caption{Cumulative utility as a function of the deviation of each individual decision variable from its optimal point, while all other variables are kept at the equilibrium. The color from cold to warm corresponds to time step from low $(t=0)$ to high $(t=39)$.}
\label{verify_betaNash}
\end{figure}

To verify the solutions actually satisfy the equilibrium condition Eq.(\ref{NashEquilibrium}), we plot the cumulative utility as a function of the deviation for individual decision variables $(\alpha_{i,t}^*,\delta_{i,t}^*)$ from its optimal point while keeping all other decision variables at the equilibrium, for $i\in \{1,\, 2\}$ and $t\in\{0,\,1,\cdots,\,39\}$. As can be seen from Fig.\ref{verify_betaNash}, the equilibrium condition is indeed met.

\subsubsection{Existence of Multi Equilibria}
Since {\it betaNash} is an iterative procedure, it depends on where the search starts, there is no guarantee that the best response dynamics will converge to the same fixed point, unless the game is supermodular as required by \cite{Milgrom90}. Unfortunately, the game in our setting is unlikely supermodular and the existence of multiple-equilibrium could happen, due to the complicated tradeoffs among various components in the utility function. This turns out to be the case, not too surprisingly. For example, a rear merge outcome could occasionally appear even under IC1, though much less likely than a front merge, if we start the best response dynamics iteration randomly. Numerically, we did not find additional equilibria, other than the aforementioned front and rear merges. It is possible that the best response dynamics in our setting is locally supermodular.

\subsection{Solution from {\it adaptiveSeek}}
In this case, we either regard the two vehicles as autonomous, as in the case of {\it betaNash}. Or, if we assume the parameters and weights in the utility functions as calibrated by the actual data, we can view the solutions in this subsection as describing the driving behaviors of two human-driven vehicles.
\subsubsection{Specification of the Utility Function}
The utility functions and parameters are identical with what were described in subsection \ref{UtilityTwoLaneHighway}. In order to guarantee that both vehicles see the barrier at $t=0$, we need to enlarge the risk premium parameter for crash $l_x^{(7)}=10.0\,m$, and set the anticipation horizon to $3$ seconds, implying $h=15$.

\subsubsection{State Evolution and Path Anticipation}
The self state evolution per period is the same as in Eq.(\ref{StateEvolutionBicycleModel},\ref{HeadingAngle},\ref{SpeedEvolution}) when the action is given. However, to calculate the effective utility function in Eq.(\ref{EffectiveUtilityTilde}) we need to anticipate future path scenarios. In this particular example, this amounts to inferring whether the other vehicle will change lane and how. For the lane change maneuvering, we follow a slightly modified Stanley's orientation formula \cite{Stanley06} shown below:
\begin{equation}
\psi_t^{*} = \tan^{-1}\Big(\frac{\kappa\,d_t}{\sqrt{1+v_t}}\Big)\, , \label{Stanley}
\end{equation}
where $d_t$ is the distance between the center of gravity of the vehicle and the center of the lane, $v_t$ is the velocity of the vehicle, and $\kappa=0.15$ is an empirically tuned parameter. The steering is then negatively proportional to the difference between the current orientation and the above prescribed Stanley formula, as if we were following a feedback control algorithm for the vehicle's steering. In this feedback control, Stanley's formula acts as the reference point or the guidance for setting the steering angle.  Of course, any other similar formulas will do the job for us. Stanley's formula merely provides an explicit mechanism to operationalize the path anticipation. The explicit future path anticipation during path planning is implemented by the following algorithm (Algorithm III):

\newpage
\IncMargin{1em}
\begin{algorithm}[H]
	\DontPrintSemicolon
	\SetKw{True}{true}\SetKw{Break}{break}
	\For{agent $i\in I$ with the given action $a_{i, t} = (\alpha_{i, t}, \delta_{i, t})$}{
		get current state of all agents $s'_0$ = $s_t$ for path anticipation\;
		\For{looking ahead time $\tilde{t}=1:h$}{
			\uIf{$\tilde{t}<h/3$}{
				$a'_{i, \tilde{t}} = a_{i, t}$, $a'_{-i, \tilde{t}}=(0, 0)$ \tcp*[h]{this is to maintain action persistence briefly}
			}
			\Else{
				for agent $i$:\;
				\Indp
				\SetInd{2em}{0em}
				\uIf{agent $i$ crosses the lane divider}{
					$a'_{i, \tilde{t}} = (\alpha_{i,t}, \delta'_{i,\tilde{t}}= {\psi}_{i, \tilde{t}}^{*} - \psi_{i, \tilde{t}}) $ \tcp*[h]{during lane change, steering follows Eq.(\ref{Stanley})}
				}
				\Else{
					$a'_{i, \tilde{t}} = a_{i, t} $ \tcp*[h]{maintain the intended motion if no lane-change}
				}
				\Indm
				for other agent $-i$\;
				\Indp
				\uIf{agent $-i$ crosses the lane divider:}{
					$a'_{-i, \tilde{t}} = (0, \delta'_{-i,\tilde{t}}= {\psi}_{-i, \tilde{t}}^{*} - \psi_{-i, \tilde{t}}) $ \tcp*[h]{during a lane change, the steering follows Eq.(\ref{Stanley})} 
				}
				\Else{
					$a'_{-i, \tilde{t}} = (0, 0) $ \tcp*[h]{maintain constant motion if no lane-change}
				}
				\SetInd{1em}{0em}				
			}
			update their state anticipation $s'_{\tilde{t}}$  according to Eq.(\ref{StateEvolutionBicycleModel})\;
		}
		calculate the effective utility for agent $i$ with the given action $a_{i, t}$ during the above path anticipation\;
	}
	\caption{path anticipation in {\it adaptiveSeek} at time $t$}\label{algo3}
\end{algorithm}\DecMargin{1em}

\vspace{6 mm}
\noindent It is important to emphasize again that these scenario assumptions are only used anticipatively in calculating the effective utility in Eq.(\ref{EffectiveUtilityTilde}) up to $h$-period, not for deriving the optimal action itself. The latter is done according to Eq.(\ref{BestResponseTilde}). For this reason, the precise details of the anticipation is not too critical.
 
\subsubsection{Adaptive Optimization}
\begin{figure}[!h]
\centering
\includegraphics[width=.90\textwidth]{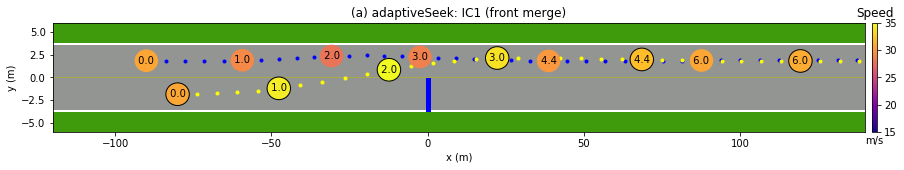}
\includegraphics[width=.90\textwidth]{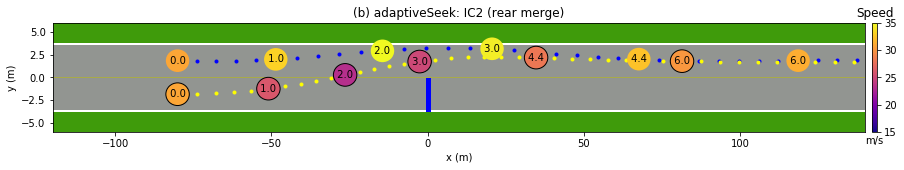}
\caption{Vehicle trajectories $(\tilde{x}_{i,t}^*,\tilde{y}_{i,t}^*), \forall t\in\{0,\, 1,\cdots 39\}$ in different merging behaviors under different initial conditions in {\it adaptiveSeek}: (a) Under IC1, the vehicle in the blocked lane (yellow dot or circle with a black frame) first accelerates longitudinally and then turns to the open lane, while the vehicle in the open lane (blue dot or square without a black frame) first yields by slowing down and then accelerates to catch up. (b) Under IC2, the vehicle in the open lane first accelerates and moves to the left slightly, while the vehicle in the blocked lane slows down first, and then turns to the open lane and accelerates to catch up. The number in the boxes represents the time in second, and the color represents the speed.}
\label{lane_change_adaptiveSeek}
\end{figure}

Without an explicit negotiation mechanism in {\it adpativeSeek}, in contrast with {\it betaNash}, the landscape for the effective utility function in Eq.(\ref{EffectiveUtilityTilde}) appears to be too complex even for {\it basinhopping} to find the right global maximum in Eq.(\ref{BestResponseTilde}). Fortunately, the action space in {\it adaptiveSeek} is small enough, we can always resort to a brute force of grid search, which guarantees a good approximation to the global optimum. The solutions found in this way are depicted in Figure.\ref{lane_change_adaptiveSeek}, with the upper panel for IC1 and lower panel for IC2. Amazingly, these solutions are nearly identical to those found by {\it betaNash} in Figure.\ref{lane_change_betaNash}, except for slightly more lateral swerving upward in the latter. 

\begin{figure}[!h]
\centering
\includegraphics[width=.85\textwidth]{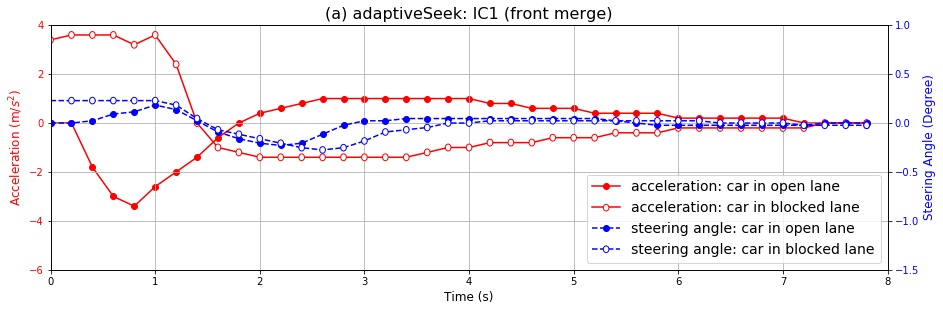} 
\includegraphics[width=.85\textwidth]{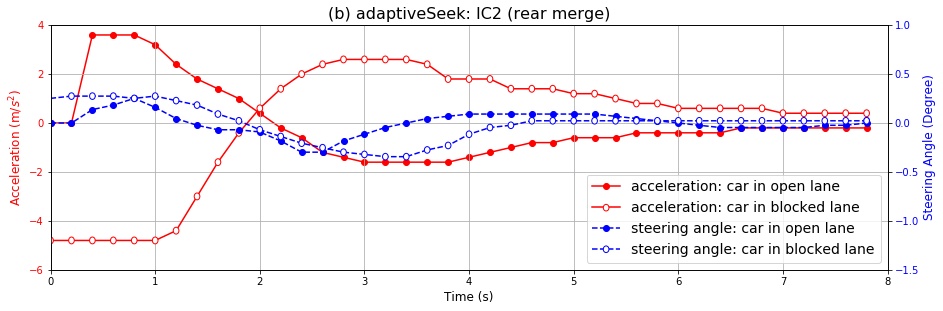} 
\caption{Optimal action sequences for both vehicles solved by {\it adaptiveSeek} under IC1 (front merge) and IC2 (rear merge).}
\label{action_sequence_adaptiveSeek}
\end{figure}

In Figure.\ref{action_sequence_adaptiveSeek} we plot the optimal action sequences for both vehicles solved from {\it adaptiveSeek} under IC1 and IC2. These sequences look very close, at least semi-quantitatively, to that derived using {\it betaNash} in Figure.\ref{action_sequence_betaNash}. Of course, there is no absolute reason to expect that the action sequences derived by using {\it betaNash} and {\it adaptiveSeek} should be quantitatively close, provided the same utility functions are used. This is because these algorithms correspond to very different solution concepts based on very different assumptions. On the other hand, it is possible that solutions from one may be able to approximate the solution of the other by slightly tweaking the utility functions or their parameters and weights. Even without doing the tweaking, the fact that the solutions already appear similar to one another is very reassuring. Furthermore, although the action sequences in Fig.\ref{action_sequence_adaptiveSeek} look a litte bit rougher than in Fig.\ref{action_sequence_betaNash}, the derived vehicle trajectories $(\tilde{x}_{i,t}^*,\tilde{y}_{i,t}^*)$ are as smooth as in the case of {\it betaNash}, because the trajectories are obtained from the action sequences by twice integration with a very small time interval. Also, a comparison between Fig.\ref{action_sequence_betaNash} and Fig.\ref{action_sequence_adaptiveSeek} shows that the merging maneuver is done (i.e. back to zero action) in about 5 to 6 seconds for {\it betaNash}, whereas at least 2 additional seconds are needed to achieve the same maneuvering in {\it adaptiveSeek}. This in turn implies that the vehicles coordinated more efficiently in the solution from {\it betaNash} than that from {\it adaptiveSeek}.

It is interesting to point out that, due to the use of grid search, {\it adaptiveSeek} is essentially a deterministic algorithm, in contrast with {\it betaNash}, given the orginal utility function, state evolution, initial condition, anticipation assumptions, and a search grid. Consequently, the situation of ``multi-equilibrium'' does not seem to arise in {\it adpativeSeek}, a  nice property at least for some cases.

\section{\bf Summary and Conclusions}
In this paper we propose a systematic computational framework for modeling smart vehicles in a smart world at micro level based on game theory. Markov games with deterministic state evolution are exploited, thanks to the explicit inclusion of risk premium in the utility functions. The corresponding sub-game perfect Nash equilibrium is solved via best response dynamics as a specific form of self-play reinforcement learning. We then relax some of the less realistic assumptions associated with Nash equilibrium and develop a heuristics based adaptive optimization method that allows us to obtain solutions that are close to the Nash equilibrium, while drastically reducing the computational burden. In this framework, inter-agent interaction is at the center of the modeling: all agents are treated equally, apart from the explicit heterogeneity in preference, intention and initial condition. We then illustrate how our approach works explicitly in a concrete example of two vehicles in a setting of a double-lane highway with an unexpected barrier. The front merge and rear merge behaviors endogenized by {\it betaNash} and {\it adaptiveSeek} are shown to be very similar to each other and both appear reasonable and intuitive. Finally, our solutions appear to be reasonably robust against minor disturbances, either in model parameters/hyperparameters and utility weights, or in adding small random noises.

So far, we have been concentrating on the forward problem exclusively in this paper. Even along this line there are many applications can be pursued immediately. The specific simulation experiment in Section \ref{SimulationExperiment} is essentially a mandatary lane change problem, deliberately chosen to be relatively simple so that all the details can be illustrated thoroughly. But our framework is much more powerful and broader, including the inverse problem and mechanism design problem mentioned in the Introduction. The latter problems will be tackled and illustrated in our forthcoming work. For example, we show in \cite{CalibrationPaper} how to use traffic video data from Sugiyama experiment \cite{Sugiyama08} for behavioral calibration with explicit heterogeneity, using {\it adaptiveSeek} as the decision-making model for human drivers. Extending this line of effort, we have started to collect naturalistic driving data using drone at a urban roundabout, and the calibration methodology outlined is being applied \cite{RoundaboutCalibration}. We demonstrate the flexibility of handling different types of agents in our approach by considering a two-way traffic with a signed or unsigned crosswalk \cite{PedestrianPaper}, where explicit interactions among vehicles, pedestrians, and stop signs are modeled. We further show in two other separate papers how game theory based coordination can be used to improve traffic flow at a single-lane roundabout \cite{CoordinationPaper}, and how smart algorithms for a few system-controlled CAVs can be invoked to tame optimally stop-and-go shockwaves \cite{CAVandShockwaves}, taking the full advantage of connectivity/autonomy and smart infrastructure.  With {\it adaptiveSeek} being able to serve as the micro path planning algorithm, we can introduce the concept of an edge-centric automated traffic system for drive-by-wire vehicles, co-mingling with human driven vehicles. This idea will be made explicit in a coordinated autonomous valet parking facility \cite{AutoValetParking} that is much more efficient in avoiding gridlock in congested places.


\section*{\bf Acknowledgement}
We thank Gint Puskorius and Jinhong Wang for several useful discussions and for their comments on the manuscript. We are also grateful to Paul Stieg for his help in literature review.

\bibliographystyle{unsrt}
\bibliography{refs}

\end{document}